\documentclass[prb,aps,twocolumn,showpacs,eqsecnum,superscriptaddress]{revtex4}
\usepackage{amssymb}
\usepackage{amsmath}
\usepackage{graphicx}
\usepackage{wasysym}
\usepackage{color}

\newcommand{\be}{\begin{eqnarray}}
\newcommand{\ee}{\end{eqnarray}}
\newcommand{\bbm}{\begin{bmatrix}}
\newcommand{\ebm}{\end{bmatrix}}
\renewcommand{\v}[1]{{\bf #1}}
\newcommand{\nn}{\nonumber \\}
\newcommand{\lb}{l_{\mathrm{B}}}

\begin{document}
\title[]{Quantum oscillations in the Luttinger model with quadratic band touching: applications to pyrochlore iridates}
%\author{Gil Dong \surname{Hong}} \email{jkps@kps.or.kr}
%\thanks{Fax: +82-2-554-1643}

\author{Jun-Won \surname{Rhim}}
\affiliation{School of Physics, Korea Institute for Advanced Study, Seoul 130-722, Korea}

\author{Yong Baek \surname{Kim}}
\affiliation{Department of Physics,
University of Toronto, Toronto, Ontario M5S 1A7, Canada}
\affiliation{Canadian Institute for Advanced Research, Toronto, Ontario, M5G 1Z8, Canada}

%\date{\today}

\begin{abstract}
Motivated by recent experiments on Pr$_2$Ir$_2$O$_7$, we provide a theory of quantum oscillations in the Luttinger model with quadratic band touching, modelled for the spin-orbit-coupled conduction electrons in pyrochlore iridates. The magneto- and Hall resistivities are computed for electron- and hole-doped systems, and the corresponding  Shubnikov de Haas (SdH) signals are investigated. The SdH signals are characterized by aperiodic behaviors that originate from the unconventional Landau level structures of the Luttinger model near the neutrality point, such as the inter-Landau level crossing, nonuniform Landau level spacings and non-parabolic dispersions along the applied magnetic-field direction. The aperiodic SdH signals observed in the paramagnetic state of Pr$_2$Ir$_2$O$_7$ are shown to be consistent with such behaviors, justifying the use of the Luttinger model and the quadratic band touching spectrum as excellent starting points for physics of pyrochlore iridates. The implications of these results are discussed in light of recent theoretical and experimental developments in these systems.
\end{abstract}

%\pacs{73.21.Ac, 73.90.tf, 73.21.-b}

\keywords{}

\maketitle

%%%%%%%%%%%%%%%%%%%%%%%%%%%%%%%%%%%%%%%%%%%%%%%%%%%%%%%%%%%%%%%%%%%%%%%%%%%%%%%%%%%%%%
%%%%%%%%%%%%%%%%%%%%%%%%%%%%%%%%%%%%%%%%%%%%%%%%%%%%%%%%%%%%%%%%%%%%%%%%%%%%%%%%%%%%%%

\section{Introduction}
Recent research activities in materials with strong spin-orbit coupling (SOC) are motivated by the prospect of discovering novel quantum ground states of interacting electrons.\cite{krempa13} It has been proposed that the combined effect of electron-electron interaction and strong spin-orbit coupling may lead to emergence of 
topological phases as well as unusual magnetically-ordered states.\cite{krempa13} 

Prominent examples are pyrochlore iridates, R$_2$Ir$_2$O$_7$, where R is a rare-earth element and Ir$^{4+}$ ions form a pyrochlore lattice.\cite{maeno,matsuhira07,matsuhira} The electronic structure of this system can be described by strongly spin-orbit-coupled $J_{\rm eff} =1/2$ degrees of freedom at Ir$^{4+}$ sites.\cite{bjkim,pesin,yang,wan1,krempa,ara,kargarian} These materials show low-temperature insulator-metal crossover when the rare-earth elements are changed, which may be due to varied relative-strength of the interaction and band-width of the itinerant electrons.\cite{maeno,matsuhira07,matsuhira} Hence they are ideal playgrounds for the interplay between the interaction and spin-orbit coupling. A number of topological phases such as topological insulator\cite{pesin,yang,kane}, axion insulator\cite{wan1}, and 
Weyl semi-metal\cite{wan1,krempa,ara}, are proposed to occur.

It has recently been shown\cite{moon} that the emergence of these novel ground states can be understood using the so-called Luttinger model\cite{luttinger} with quadratic band touching\cite{yang}as the staring point. Here the Luttinger model can be regarded as a minimal model for the Ir itinerant electrons near the Fermi level in the paramagnetic state of pyrochlore iridates. Starting from such a model, it is shown that topological insulator and Weyl semi-metal can be obtained by applying a uni-axial strain and time-reversal symmetry breaking perturbations, respectively.\cite{yang,krempa,moon} Furthermore, the effect of long range Coulomb interaction on the Luttinger model is studied, leading to
the discovery of a stable non-Fermi liquid phase.\cite{moon} Since the quadratic band touching plays such an important role, it would be important to independently verify its presence in the paramagnetic state of pyrochlore iridates.

In this paper, we study quantum oscillations in the Luttinger model with quadratic band touching. More specifically, we consider small electron- and hole-doping situations, $n_e$ or
$n_h \sim 10^{23} m^{-3}$, for the quadratic band touching. The longitudinal and Hall resistivities are computed, and the resulting SdH signals are analyzed. The main findings of theoretical analyses is that the SdH signal is aperiodic in inverse of the applied magnetic field, in contrast to usual SdH signals in systems with Fermi surfaces. This is due to unusual Landau level structures in the Luttinger model, which show the inter-Landau level crossing, nonuniform Landau level spacings and non-parabolic dispersions along the applied magnetic-field direction. The relations between the SdH signals and the structures of density of states are also pointed out. 

These theoretical results are compared to the aperiodic SdH signals observed in the paramagnetic state of Pr$_2$Ir$_2$O$_7$.\cite{machida,nakatsuji,onoda}
We conclude that slight doping of the quadratic band touching is consistent with the SdH signals measured in the experiments.
This could be taken as an independent evidence that the Luttinger model with quadratic band touching is an excellent starting point for physics of pyrochlore iridates. We notice also that a recent ARPES (angle resolved photoemission spectroscopy) study in 
the paramagnetic state of Pr$_2$Ir$_2$O$_7$ shows the quadratic band touching spectra along the [111] direction in the Brillouin zone.\cite{nakatsuji14}, which would be consistent with our results. 
While our study is motivated by physics of pyrochlore iridates, our results can be applied to transport properties of any systems described by the Luttinger model with quadratic band touching, a generic low energy model with cubic symmetry.

The rest of the paper is organized as follows. In Sec. II, we analyze the peculiar Landau level structure of the Luttinger model. After revisiting Luttinger's work briefly, we develop a perturbation theory for the transport calculations. In Sec. III, the linear response theory for the magneto-transport {\it \`{a} la} Smr$\breve{\mathrm{c}}$ka and St$\breve{\mathrm{r}}$eda is reviewed and applied to the Luttinger model.
In Sec. IV, we discuss the origins of the unconventional SdH patterns and discuss the theoretical results in the context of experimental data. 
Finally, the concluding remarks are placed in Sec. V.

%%%%%%%%%%%%%%%%%%%%%%%%%%%%%%%%%%%%%%%%%%%%%%%%%%%%%%%%%%%%%%%%%%%%%%%%%%%%%%%%%%%%%%
%%%%%%%%%%%%%%%%%%%%%%%%%%%%%%%%%%%%%%%%%%%%%%%%%%%%%%%%%%%%%%%%%%%%%%%%%%%%%%%%%%%%%%

\section{Landau levels of the Luttinger model}

%%%%%%%%%%%%%%%%%%%%%%%%%%%%%%%%%%%%%%%%%%%%%%%%%%%%%%%%%%
\begin{figure*}
\includegraphics[width=2\columnwidth]{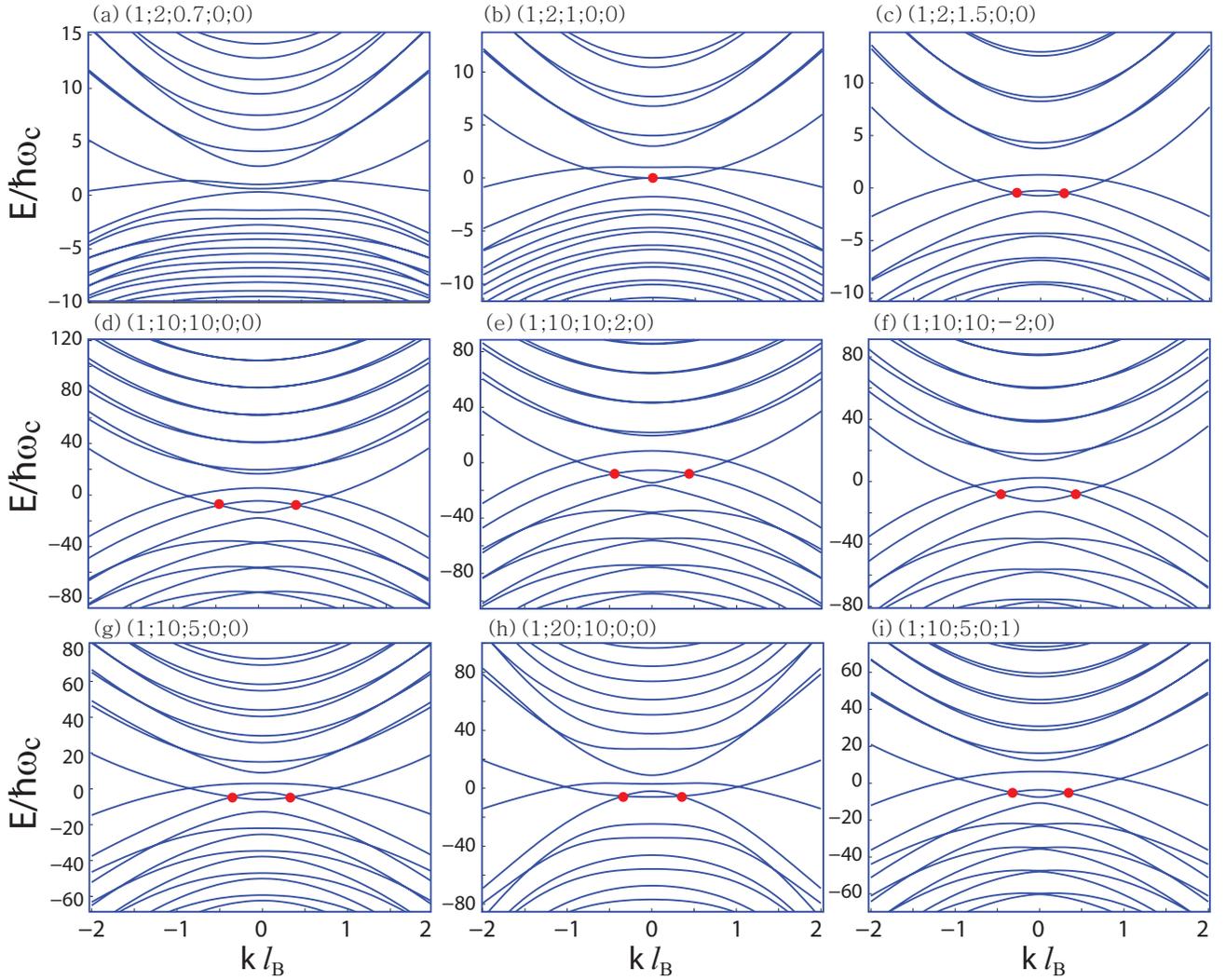}
\caption{(Color online) Landau level dispersions of the Luttinger model near the quadratic band touching point with various band parameters $(\gamma_1;\gamma_2;\gamma_3;\kappa;q)$. The red dots represent the band crossings between two consecutive low-lying Landau levels. As shown in (a), (b), (c), the Landau level crossings appear when $\gamma_3 > \gamma_1$.}
\label{fig:band}
\end{figure*}
%%%%%%%%%%%%%%%%%%%%%%%%%%%%%%%%%%%%%%%%%%%%%%%%%%%%%%%%%%

\subsection{General case}
If the magnetic field is applied to a system possessing the four-fold degeneracy at the zone center ($\v k = 0$) and the cubic symmetry, its low energy physics can be described by the Luttinger Hamiltonian\cite{luttinger}
\be
H_L &=& \frac{\hbar^2}{m} \bigg\{ \left( \gamma_1 + \frac{5\gamma_2}{2}\right) \frac{k^2}{2} - \gamma_2 (k_x^2 J_x^2+k_y^2 J_y^2+k_z^2 J_z^2) \nn
&& -2\gamma_3 \Big( \{k_x,k_y\}\{J_x,J_y\}+\{k_y,k_z\}\{J_y,J_z\} \nn
&& +\{k_z,k_x\}\{J_z,J_x\} \Big) +\frac{e}{c}\kappa \v J\cdot\v B  + \frac{e}{c}q \sum_\alpha B_\alpha J_\alpha^3 \bigg\}, \nn \label{eq:luttinger0}
\ee
where $\{a,b\}=(ab+ba)/2$, $\v B$ is the applied magnetic field and $J_{\alpha}$ ($\alpha = x, y, z$) are the $J=3/2$ angular momentum operators.
Here, the angular momentum operators are used as bases for the four bands under consideration.
The dimensionless Luttinger parameters $\gamma_1$, $\gamma_2$ and $\gamma_3$, control the effective masses of four conduction and valence bands.
Without the magnetic field, the quadratic band touching between the conduction and valence bands is realized when $\gamma_3 > \gamma_1/2$. \cite{moon}
While $\kappa$ and $q$ are also dimensionless, $\kappa$ can be regarded as the coefficient of the effective total magnetic moment and $q$ is a parameter related to the SOC.
In many cases, the magnitude of $q$ is much smaller than $\gamma_i$.\cite{phillips, klipstein, yablonovitch} For example, $q/\gamma_1$ is about 0.01 in GaAs. Since the SOC strength of Ir atom is one order of magnitude greater than that of Ga or As, it is reasonable to expect that the value of $q/\gamma_i$ of iridates would be order of 0.1.
The kinetic momentum operators $k_\alpha = p_\alpha +(e/c)A_\alpha$, where $\v A$ is the vector potential, do not commute with each other.
Instead, they satisfy the commutation realation $[k_\alpha,k_\beta] = -i\epsilon_{\alpha\beta\gamma}(e/c)B_\gamma$, where $\epsilon_{\alpha\beta\gamma}$ is the totally antisymmetric tensor.

Considering the magnetic field along $[111]$ direction, the natural choice of the coordinates $k_1$, $k_2$ and $k_3$ are given by the transformations $k_x=(1/\sqrt{6})k_1-(1/\sqrt{2})k_2+1/(\sqrt{3})k_3$, $k_y=(1/\sqrt{6})k_1+(1/\sqrt{2})k_2+(1/\sqrt{3})k_3$ and $k_z=-(\sqrt{2/3})k_1+(1/\sqrt{3})k_3$.
They can be rewritten as $k_1 = (eB/2c)^{1/2}(a^\dag + a)$ and $k_2= -i(eB/2c)^{1/2}(a^\dag - a)$ by using the ladder operators $a^\dag$ and $a$.

While the Luttinger Hamiltonian can be written in terms of a $4\times 4$ matrix representation, it is difficult to find the solutions for general cases
because its matrix elements are not numbers, but operators. Luttinger showed that his model can be analytically solved only for some special cases such as (a) $k_3=0$ with $\v B \parallel [111]$ and (b) $q=0$ and $\gamma_2=\gamma_3$. Here, $k_3$ is the momentum component along the field direction. In those cases, one can find a simple form of the eigenvector involving finite number of the harmonic oscillator eigenfunctions. As a result, the Hamiltonian has a $4\times 4$ matrix representation composed only of numbers as elements. These cases are discussed in detail in Sec. II. B.

Here we consider generic cases and present a numerical solution. We start from the eigenvector with an infinite sum of the  Harmonic oscillator eigenfunctions $u_n$:
\begin{eqnarray}
|\psi\rangle = \sum_{n=0}^\infty \bbm C^{(1)}_{n,k_3} u_n & C^{(2)}_{n,k_3} u_n & C^{(3)}_{n,k_3} u_n & C^{(4)}_{n,k_3} u_n\ebm^{\mathrm{T}}.
\end{eqnarray}
Then, the eigenvalue problem $H_L |\psi\rangle = \varepsilon |\psi\rangle$ leads to the following infinite number of coupled secular equations:
\begin{widetext}
\be
(\varepsilon - h^{(1)}_{n,k_3}) C^{(1)}_{n,k_3} &=& -g_1\sqrt{(n+2)(n+1)} C^{(2)}_{n+2,k_3} + s_1\sqrt{n} \bar{k}_3 C^{(2)}_{n-1,k_3} - g_2\sqrt{n(n-1)} C^{(3)}_{n-2,k_3} \nn 
&& - s_2\sqrt{n+1}\bar{k}_3 C^{(3)}_{n+1,k_3} -\frac{q}{\sqrt{2}} C^{(4)}_{n,k_3} \\
(\varepsilon - h^{(2)}_{n,k_3}) C^{(2)}_{n,k_3} &=& -g_1\sqrt{n(n-1)} C^{(1)}_{n-2,k_3} +s_1\sqrt{n+1}\bar{k}_3 C^{(1)}_{n+1,k_3} + g_2\sqrt{n(n-1)} C^{(4)}_{n-2,k_3} \nn
&& + s_2\sqrt{n+1}\bar{k}_3 C^{(4)}_{n+1,k_3} \\
(\varepsilon - h^{(3)}_{n,k_3}) C^{(3)}_{n,k_3} &=& -g_2\sqrt{(n+2)(n+1)} C^{(1)}_{n+2,k_3} -s_2\sqrt{n}\bar{k}_3 C^{(1)}_{n-1,k_3} - g_1\sqrt{(n+2)(n+1)} C^{(4)}_{n+2,k_3} \nn
&& + s_1\sqrt{n}\bar{k}_3 C^{(4)}_{n-1,k_3} \\
(\varepsilon - h^{(4)}_{n,k_3}) C^{(4)}_{n,k_3} &=& -\frac{q}{\sqrt{2}}C^{(1)}_{n,k_3} + g_2\sqrt{(n+2)(n+1)} C^{(2)}_{n+2,k_3} + s_2\sqrt{n}\bar{k}_3 C^{(2)}_{n-1,k_3} \nn
&& -g_2\sqrt{n(n-1)}C^{(3)}_{n-2,k_3} +s_1\sqrt{n+1}\bar{k}_3 C^{(3)}_{n+1,k_3}
\ee
\end{widetext}
where
\be
h^{(1)}_{n,k_3} &=& (\gamma_1+\gamma_3)\frac{2n+1}{2}+\frac{3}{8}\kappa+\frac{23}{8}q+\frac{(\gamma_1-2\gamma_3)}{2}\bar{k}_3^2, \nn
h^{(2)}_{n,k_3} &=& (\gamma_1-\gamma_3)\frac{2n+1}{2}-\frac{1}{2}\kappa-\frac{13}{8}q+\frac{(\gamma_1+2\gamma_3)}{2}\bar{k}_3^2, \nn
h^{(3)}_{n,k_3} &=& (\gamma_1-\gamma_3)\frac{2n+1}{2}+\frac{1}{2}\kappa+\frac{13}{8}q+\frac{(\gamma_1+2\gamma_3)}{2}\bar{k}_3^2, \nn
h^{(4)}_{n,k_3} &=& (\gamma_1+\gamma_3)\frac{2n+1}{2}-\frac{3}{2}\kappa-\frac{23}{8}q+\frac{(\gamma_1-2\gamma_3)}{2}\bar{k}_3^2, \nn
\ee
$g_1=(\gamma_2+2\gamma_3)/\sqrt{3}$, $g_2=-\sqrt{2/3}(\gamma_2-\gamma_3)$, $s_1=2(\gamma_2-\gamma_3)/\sqrt{3}$ and $s_2=\sqrt{2/3}(2\gamma_2+\gamma_3)$.
The coefficient $C^{(m)}_n$ vanishes when $n$ is negative.

While this set-up allows us to study the Luttinger model non-perturbatively for arbitrary Luttinger parameters and momenta, finding the infinite number of coefficients $C^{(m)}_n$ is an extremely difficult task.
However, if we are interested in the spectra when the band-filling is close to the half-filling, this problem can be resolved by setting a cutoff for the sum over $n$.
The solutions of the Luttinger model for the case (a) and (b) indicate that the eigenvectors near the band touching point involve only small quantum numbers $n<n_c$.
Even if $k_3$ is nonzero and $\gamma_2 \neq \gamma_3$, the contributions from the higher Landau levels ($ n \gg n_c$) to the low-lying Landau levels ($ n < n_c$) are expected to be small and we checked that they are indeed negligible even for large momenta.
Given a large enough cutoff $n=N \gg n_c$, the coefficient $C^{(m)}_n$ is assumed to be zero when $n>N$ so that the secular equations become an eigenvalue problem of a $4N\times 4N$ matrix with elements provided by the above secular equations.
This process gives us reliable solutions for the Landau levels involving $u_n$'s with $n<n_c$.

In Fig. 1, we plot energy spectra of the Luttinger Hamiltonian for wide range of the Luttinger parameters satisfying two constraints (i) $\gamma_3 > \gamma_1/2$ (quadratic band touching) and (ii) $\gamma_2 \geq \gamma_3$.
The latter condition reflects the fact that the effective mass along $\Gamma$-L is heavier than that along $\Gamma$-K in the previous tight binding model calculations.
From these energy spectra, we note several generic properties of the Landau levels of the Luttinger model with quadratic band touching.
First, the Landau level crossing away from the half-filling appears for a wide range of the Luttinger parameters when $\gamma_1 < \gamma_3$ as represented by red dots.
This happens because the minimum of the $n=-1$ conduction Landau level is $\frac{3}{2}(\gamma_1-\gamma_3) -\kappa/2-13q/8$ while the maximum of the $n=-2$ valence Landau level is $\frac{1}{2}(\gamma_1-\gamma_3) -\kappa/2-13q/8$ (Notice that, in this discussion, we are using the Luttinger's convention for the Landau level labelling for the case of $k_3=0$ in his work\cite{luttinger}). Hence, the level crossing between $n=-1$ and $n=-2$ Landau levels are inevitable when $\gamma_1 - \gamma_3$ is negative.
Second, the non-uniform Landau level spacings and effective masses in the Landau level structures near the quadratic band touching point are ubiquitous for any set of the Luttinger parameters.

%%%%%%%%%%%%%%%%%%%%%%%%%%%%%%%%%%%%%%%%%%%%%%%%%%%%%%%%%%
\begin{figure*}
\includegraphics[width=2\columnwidth]{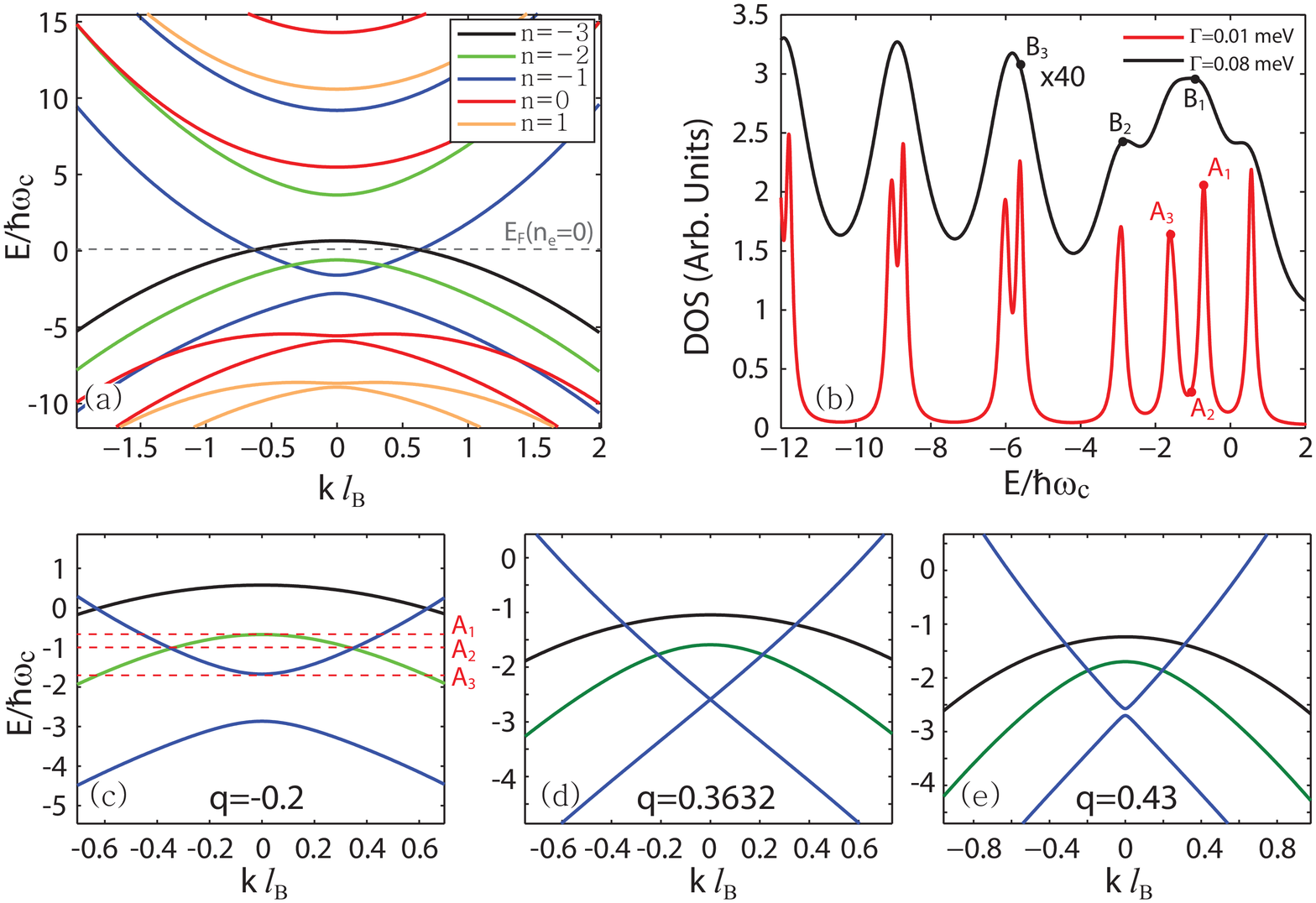}
\caption{(Color online) (a) Landau level spectra for a set of Luttinger parameters $(\gamma_1;\bar{\gamma};\kappa;q)=(1;2;1;-0.2)$ as functions of $k_3 \lb$. The dashed line near the zero energy represents the Fermi level at half-filling. (b) Densities of states for the dispersions in (a) with Landau level broadenings $\Gamma=0.01$ meV and $\Gamma = 0.08$ meV. For both cases, the strength of the magnetic field is $5$T. Panels from (c) to (e) exhibit the SOC-induced band inversion for $n=-1$ Landau levels. Once they experience the gap-closing, as $q$ grows, the band gap starts to increase again.}
\label{fig:band}
\end{figure*}
%%%%%%%%%%%%%%%%%%%%%%%%%%%%%%%%%%%%%%%%%%%%%%%%%%%%%%%%%%

\subsection{Isotropic case $\gamma_2 = \gamma_3$}

Let us now consider the special case of $\gamma_2 = \gamma_3$ or the case (b) mentioned earlier. 
The advantage of this case is that one can obtain an analytic solution, which becomes useful when we consider transport properties in the Littinger model.
It is shown below that general features of the spectra in this case are qualitatively similar to those in the general cases considered in the previous section.
Here, the Luttinger Hamiltonian is reduced to a simpler form:
\be
H^{b}_L =\frac{\hbar^2}{m} \bigg\{ \gamma_{12}\frac{k^2}{2} - \bar{\gamma} \left(\v k\cdot\v J \right)^2 +(\kappa-\frac{\bar{\gamma}}{2})\frac{e}{c}\v J\cdot \v B\bigg\},\label{eq:luttinger1}
\ee
where $\bar{\gamma} = \gamma_2 = \gamma_3$ and $\gamma_{12}=\gamma_1+(5/2)\bar{\gamma}$.
The last term of Eq.(\ref{eq:luttinger0}), $H_q = \frac{e}{c}q \sum_\alpha B_\alpha J_\alpha^3$, would be regarded as a perturbation.
While the reduced Hamiltonian (\ref{eq:luttinger1}) is rotationally invariant so that the Landau level structures are independent of the field direction, we assume that $\v B$ is along [111] direction in the rest of the paper for convenience. Notice, however, that $H_q$ term would break the rotational invariance and would make the magnetic response anisotropic.
If we adopt the Luttinger's choice for the representations of the angular momentum operators\cite{luttinger}, the matrix form of the Hamiltonian (\ref{eq:luttinger1}) is given by
\begin{widetext}
\be
H^b_L =\hbar\omega_c \bbm d_1\hat{N}+\dfrac{d_1+3\kappa}{2}+d_3\bar{k}_3^2 & -\sqrt{3}\bar{\gamma}a^2 & -\sqrt{6}\bar{\gamma}\bar{k}_3a & 0 \\ -\sqrt{3}\bar{\gamma}a^{\dag 2} & d_2\hat{N}+\dfrac{d_2-\kappa}{2}+d_4\bar{k}_3^2 & 0 & \sqrt{6}\bar{\gamma}\bar{k}_3a \\ -\sqrt{6}\bar{\gamma}\bar{k}_3a^\dag & 0 & d_2\hat{N}+\dfrac{d_2+\kappa}{2}+d_4\bar{k}_3^2 & -\sqrt{3}\bar{\gamma}a^{2} \\ 0 & \sqrt{6}\bar{\gamma}\bar{k}_3a^\dag & -\sqrt{3}\bar{\gamma} a^{\dag 2} & d_1\hat{N}+\dfrac{d_1-3\kappa}{2}+d_3\bar{k}_3^2\ebm \label{eq:hamiltonian_matrix},
\ee
\end{widetext}
where $d_1=\gamma_1+\bar{\gamma}$, $d_2=\gamma_1-\bar{\gamma}$, $d_3=(\gamma_1-2\bar{\gamma})/2$, $d_4=(\gamma_1+2\bar{\gamma})/2$, $\hat{N}=a^\dag a$, $\bar{k}_3 = \lb k_3$ and the magnetic length $\lb = \sqrt{\hbar/eB}$.
In the same way, the perturbative part has its matrix representation as
\be
H_q = \hbar\omega_c q\bbm \frac{23}{8} & 0 & 0 & -\frac{1}{\sqrt{2}} \\ 0 & -\frac{13}{8} & 0 & 0 \\ 0 & 0 & \frac{13}{8} & 0 \\ -\frac{1}{\sqrt{2}} & 0 & 0 & -\frac{23}{8}\ebm.
\ee

The unperturbed model can be solved by a trial state
\be
\psi^{n,j}_{k_3} = \bbm c^{n,j}_{k_3,1} u_n & c^{n,j}_{k_3,2} u_{n+2} & c^{n,j}_{k_3,3} u_{n+1} & c^{n,j}_{k_3,4} u_{n+3} \ebm^{\mathrm{T}}
\ee
where $u_n$ is the coherent state of the harmonic oscillator for $n \ge 0$ and zero for $n<0$.
While the integer number $n$ runs from $-3$, the other integer $j$ in $c^{n,j}_{k_3}$ can have its value from 1 to $n+4$ for $n<0$ and from 1 to 4 for $n \ge 0$.
By solving the secular equation for the $4\times 4$ matrix constructed by the coherent states, one can obtain the eigenenergies $\epsilon^{n,j}_{k_3}$ and the coefficients $c^{n,j}_{k_3,i}$ as eigenvectors.

%When $k_3=0$, $H^b_L$ is separable and analytic solutions are allowed as follows.
%\be
%\epsilon^{-3,1}_0 &=& \frac{1}{2}(\gamma_1+\bar{\gamma})-\frac{3}{2}\kappa, \\
%
%\epsilon^{-2,j}_0 &=& \frac{1}{2}(\gamma_1-\bar{\gamma})-\frac{1}{2}\kappa \quad\mathrm{and}\quad \frac{3}{2}(\gamma_1+\bar{\gamma})-\frac{3}{2}\kappa, \quad\quad \\
%
%\epsilon^{-1,j}_0 &=& \frac{3}{2}(\gamma_1-\bar{\gamma})-\frac{1}{2}\kappa \quad\mathrm{and}\quad \varepsilon^\pm_2(2),
%\ee
%and, for the case of $n \ge 0$,
%\be
%\epsilon^{n,j}_0 &=& \varepsilon^\pm_1(n+2) \quad\mathrm{and}\quad \varepsilon^\pm_2(n+3)
%\ee
%where $\varepsilon^\pm_1(n) = \gamma_1 n - (\gamma_1/2+\bar{\gamma}-\kappa/2) \pm [ \{\bar{\gamma}n-(\gamma_1-\kappa+\bar{\gamma}/2)\}^2+3\bar{\gamma}^2n(n-1) ]^{1/2}$ and $\varepsilon^\pm_2(n) = \gamma_1 n - (\gamma_1/2-\bar{\gamma}+\kappa/2) \pm [ \{\bar{\gamma}n+(\gamma_1-\kappa-\bar{\gamma}/2)\}^2+3\bar{\gamma}^2n(n-1) ]^{1/2}$.

The perturbative correction of $H_q$ can be computed from its matrix element as
\be
\frac{\langle \psi_{k_3}^{m,i} | H_q | \psi_{k_3}^{n,j} \rangle}{\hbar \omega_c q} &=& \delta_{m,n}\bigg\{ \frac{23}{8}\left(c^{m,i}_1c^{m,j}_1 -c^{m,i}_4c^{m,j}_4\right) \nn 
&&-\frac{13}{8}\left( c^{m,i}_2c^{m,j}_2 -c^{m,i}_3c^{m,j}_3\right)\bigg\}\nn
&&-\delta_{m,n-3}\frac{1}{\sqrt{2}}c^{m,i}_4c^{m+3,j}_1 \nn
&&-\delta_{m,n+3}\frac{1}{\sqrt{2}}c^{m,i}_1c^{m-3,j}_4,
\ee
where $c^{m,i}_1 \equiv c^{m,i}_{k_3,1}$ etc.
From this, one can note a selection rule for $H_q$; the states on the $n$-th Landau levels have nonzero matrix elements only with those on the $n-3$, $n$ and $(n+3)$-th Landau levels.
As shown in Fig. \ref{fig:band}(a), the energy difference between $n$-th and $n-3$ or $(n+3)$-th Landau levels is usually nonzero and much larger than $\hbar\omega_c$.
As a result, for $q\ll 1$, we can only consider the first order contributions within the degenerate perturbation scheme, where the dimension of the Hilbert space is reduced to the number of the n-th Landau levels.

As an example of the quadratic band touching case, we consider a set of Luttinger parameters, $\gamma_1=1$, $\bar{\gamma}=2$, $\kappa=1$ and $q=-0.2$.
The Landau levels for this parameter set are drawn in Fig. \ref{fig:band}(a) as functions of $k_3\lb$.
Unlike the usual Landau level structures for the quadratic band, the Landau level spectra for the Luttinger model with the quadratic band touching display many unconventional features such as inter-Landau level crossings, nonuniform Landau level spacing and various effective masses.
Since all the field dependences are included in $k_3\lb$ after factoring out the cyclotron energy of (\ref{eq:hamiltonian_matrix}), the Landau level structures in Fig. \ref{fig:band}(a) are independent of the magnetic field.

Although only the electronic structures near the zone center can be obtained by the Luttinger model, one can still determine the Fermi level at half filling 
by considering the finite number of Landau levels when the number of the conduction and valence bands are equal to each other.
In the extreme case, when the Landau level broadening is vanishing ($\Gamma=0$), the Fermi level at half-filling is just the energy at the crossing point between $n=-3$ and $-1$-th Landau levels as indicated by the grey dashed line in Fig. \ref{fig:band}(a).
If $\Gamma >0$, we increase the cutoff of the Landau level index $n$ until the Fermi levels obtained from the finite number of Landau levels are unaffected by further increment of the cutoff.
In the limit of vanishing magnetic field, we would recover the quadratic band-touching at the Fermi level because the distance $\Delta k_3$ between those two crossing points is proportional to $1/\lb$.
The transformation of the quadratic band touching into the band crossing is one of the nontrivial quantum effects which might be ascribed to the multi-orbital nature of the model system as well as the Zeeman splitting.

We note another interesting feature of the Luttinger's Landau levels; the inversion between Landau levels occurs as $q$ varies as shown in Fig. \ref{fig:band} (c) to (e) for the $n=-1$ Landau levels.
Those two eigen-energies of $n=-1$ Landau levels at $k_3=0$ are evaluated as $E^{-1,1}_{0}=-2-(13/8)q$ for the conduction band and $E^{-1,2}_{0}=3-(5/8)q-(81q^2-216q+528)^{1/2}/4$ for the valence band.
They move to each other in energy as a function of $q$ and the gap closes when $q=4(47-\sqrt{1689})/65\approx 0.3632$.
Further increase in $q$ results in the re-opening of the band gap.
This band inversion is also observed in other valence Landau levels.
During this process, the coefficients of the eigenvectors are relatively insensitive to the variation of $q$ (less than 10\% of the unperturbative ones up to $q=0.5$).
We call this behavior as the SOC-induced Landau level inversion since $q$ is controlled by the SOC.
% and vanishing when the SOC is absent.
In the rest of the paper, for simplicity we focus on the $\gamma_2=\gamma_3$ case that captures all the qualitative features of the Landau level spectra in general cases.

%%%%%%%%%%%%%%%%%%%%%%%%%%%%%%%%%%%%%%%%%%%%%%%%%%%%%%%%%%%%%%%%%%%%%%%%%%%%%%%%%%%%%%
%%%%%%%%%%%%%%%%%%%%%%%%%%%%%%%%%%%%%%%%%%%%%%%%%%%%%%%%%%%%%%%%%%%%%%%%%%%%%%%%%%%%%%

\section{magneto-transport in the Luttinger model}

The zero-temperature magneto-conductivity $\sigma_{11}$ and the Hall conductivity $\sigma_{12}$ are calculated by applying the formula constructed by Smr$\breve{\mathrm{c}}$ka and St$\breve{\mathrm{r}}$eda\cite{smrcka-streda1,smrcka-streda2,streda}, which is given by
\begin{widetext}
\be
\sigma_{11} &=& i\frac{\hbar e^2 }{V} \int_{-\infty}^{\mu} d\eta \left\langle \mathrm{Tr}\left\{ \delta(\eta-\hat{H})v_x\frac{\partial G^+}{\partial \eta}v_x - \delta(\eta-\hat{H})v_x\frac{\partial G^-}{\partial \eta}v_x \right\} \right\rangle \\
\sigma_{12} &=& i\frac{1}{2}\frac{\hbar e^2}{V} \left\langle \mathrm{Tr}\left\{ v_x G^+(\mu)v_y\delta(\mu-\hat{H}) -  v_y G^-(\mu)v_x\delta(\mu-\hat{H})\right\}\right\rangle -e\frac{\partial N(\mu)}{\partial B}
\ee
\end{widetext}
where $V$ is the system's volume, $G^\pm = (\eta - \hat{H} \pm i\epsilon)^{-1}$, and $N(\mu)$ is the density of electrons below the Fermi energy.
The average is taken over the impurities and the averaged Green's function is expressed as $\langle G(z) \rangle = (z - \hat{H}_0 +i\Gamma)^{-1}$, where
$\Gamma$ represents the impurity scattering.\cite{isihara,velicky} 
Instead of applying the self-consistent scheme\cite{soven}, we simply use a constant level-broadening parameter and neglect possible level shifts 
as we concern mostly about qualitative features of the transport properties.\cite{endo}
We believe that the main conclusions of our simple model would be robust in the weak impurity scattering limit.
If the impurity scattering strength is weak enough, the overlaps between broadened Landau levels are negligible.
In this case, the complex self-energy structures such as the level-dependent Landau level broadenings and shifts which are expected from an anisotropic model would cause only minor effects in the periodicity of the SdH signals.
Now the conductivities can be written as
\begin{widetext}
\be
\sigma_{11} &=& \frac{\hbar e^2}{4\pi^3 \lb^2} \sum_{n,j,m,i} \int dk_3  \left| \langle \psi_{k_3}^{n,j} | v_1| \psi_{k_3}^{m,i}\rangle \right|^2 \frac{\Gamma}{\left(\mu - E_{k_3}^{n,j}\right)^2 +\Gamma^2} \frac{\Gamma}{\left(\mu - E_{k_3}^{m,i}\right)^2 +\Gamma^2} \label{eq:cond11} \\
\sigma_{12} &=& -\frac{\hbar e^2 }{4\pi^3 \lb^2} \sum_{n,j,m,i} \int dk_3 \mathrm{Im}\left[  \frac{\langle \psi_{k_3}^{n,j} | v_1 |\psi_{k_3}^{m,i} \rangle \langle \psi_{k_3}^{m,i}| v_2 |\psi_{k_3}^{n,j} \rangle}{\mu - E_{k_3}^{m,i}+i\Gamma } \right]\frac{\Gamma}{\left(\mu - E_{k_3}^{n,j}\right)^2 +\Gamma^2 } -e\frac{\partial \Delta N(\mu)}{\partial B},
\ee
\end{widetext}
where $E_{k_3}^{m,i}$ is the eigenvalue including the perturbative correction via $H^b_L+H_q$.
Here, $\Delta N(\mu) = N(\mu) -N_0$ is the density of electrons measured from the half-filled case ($N_0$) which, as a constant, does not affect the derivative $\partial N/\partial B$.
Since we assume cutoffs for $n$ and $k_3$ in real evaluations, this relative quantity is more natural and would yield no difference in calculating $\partial N/\partial B$, compared with the full-band consideration of the lattice model if we deal with small enough carrier densities and large enough magnetic fields.
The relative density is represented as
\be
\Delta N(\mu) = \frac{1}{4\pi^3\lb^2}\sum_{n,j}\int dk_3 \left( \tan^{-1}\frac{\mu-E_{k_3}^{n,j}}{\Gamma} +\frac{\pi}{2}\right). \quad
\ee

%%%%%%%%%%%%%%%%%%%%%%%%%%%%%%%%%%%%%%%%%%%%%%%%%%%%%%%%%%
\begin{figure*}
\includegraphics[width=2\columnwidth]{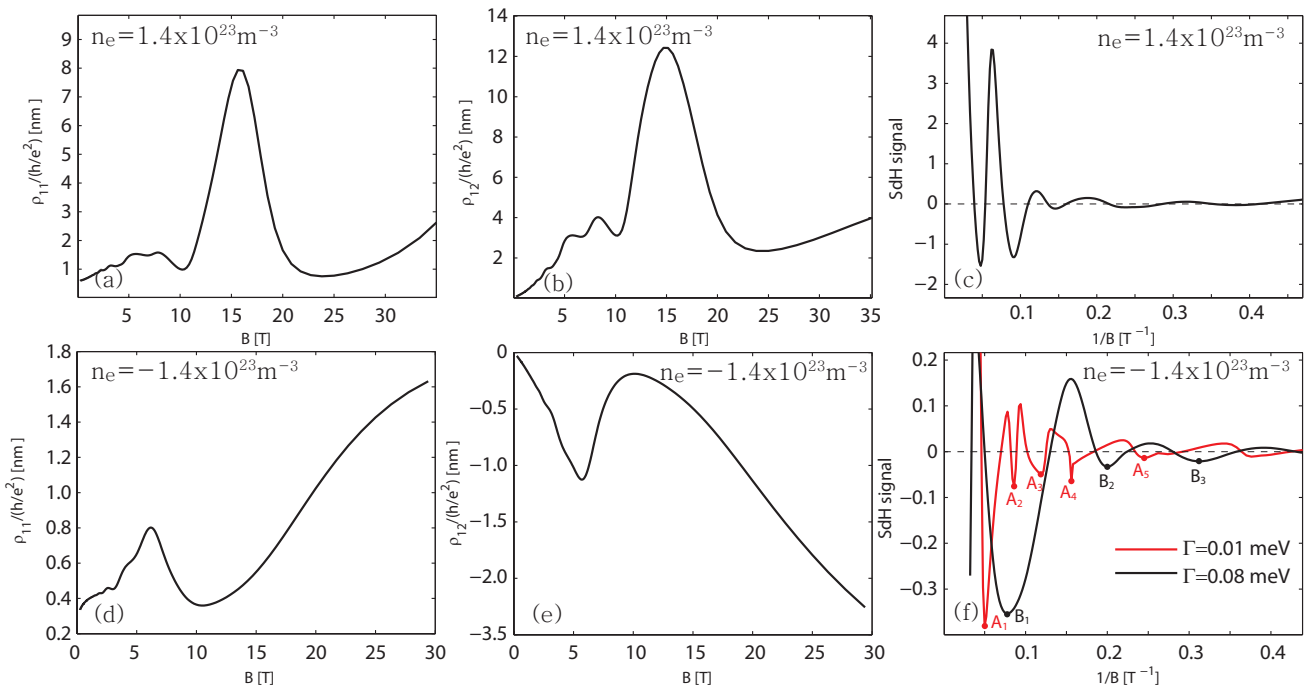}
\caption{(Color online) The magneto- and Hall resistivities, and the SdH signals are plotted for both electron and hole doped cases in the upper ((a) to (c)) and lower ((d) to (f)) panels, respectively. In (f), the SdH signals with different impurity strengths are compared to each other. The indices A$_i$ and B$_i$ represent the local minima of the curves. Densities of states and the Fermi levels corresponding to some of the minima in the SdH signal are shown in Fig. \ref{fig:band}(b) and (c). }
\label{fig:transport}
\end{figure*}
%%%%%%%%%%%%%%%%%%%%%%%%%%%%%%%%%%%%%%%%%%%%%%%%%%%%%%%%%%

The velocity operator for the Luttinger model, $v_\sigma = (i/\hbar)[H,x_\sigma]$, is given by
\be
v_\sigma &=& \frac{i\hbar}{2m} \gamma_{12} [k^2,x_\sigma] - \frac{i\hbar}{m} \bar{\gamma} \sum_{\alpha,\beta} [\{k_\alpha,k_\beta\},x_\sigma] \{J_\alpha,J_\beta\} \nn
&=& \frac{\hbar}{m} \gamma_{12} k_\sigma  + \frac{2\hbar}{m}\gamma_2\sum_{\alpha} k_\alpha \{J_\alpha,J_\sigma\},
\ee
where the identity $\sum_{\alpha,\beta}[\{k_\alpha,k_\beta\},x_\sigma]\{J_\alpha,J_\beta\} = -2i\sum_{\alpha} k_\alpha \{J_\alpha,J_\sigma\}$ is used.
To understand the selection rule of the velocity operator, let us start by defining a `seed' state $|\phi_n\rangle = \bbm u_n & u_{n+2} & u_{n+1} & u_{n+3}\ebm^{\mathrm{T}}$ which has nonzero overlap only with eigenstates with the same Landau level index $n$, {\it i.e.}, $| \psi^{n,j}_{k_3}\rangle$.
Using this definition, the eigenstate can be re-expressed as $| \psi^{n,j}_{k_3}\rangle = \v C^{n,j}_{k_3} |\phi_n\rangle$, where $\v C^{n,j}_{k_3}$ is the $4\times 4$ diagonal matrix with components $[\v C^{n,j}_{k_3}]_{ii} = c^{n,j}_{k_3,i}$.
Then, a new state obtained by the operation of the velocity operator ($\sigma=1$ or 2) on $\psi^{n,j}_{k_3}$ is given by
\be
v_\sigma |\psi^{n,j}_{k_3}\rangle =\frac{(-i)^{\sigma-1}\hbar}{\sqrt{2}m\lb}\left( \v A^{n,j}_{k_3,\sigma}|\phi_{n-1}\rangle + \v B^{n,j}_{k_3,\sigma}|\phi_{n+1}\rangle \right)\quad\label{eq:selection_rule}
\ee
where the elements of the diagonal matrices $\v A^{n,j}_{k_3,\sigma}$ and $\v B^{n,j}_{k_3,\sigma}$ read that
$[\v A^{n,j}_{k_3,\sigma}]_{11}=\pm c^{n,j}_{k_3,1}\sqrt{n}(\gamma_1+4\bar{\gamma})$, $[\v A^{n,j}_{k_3,\sigma}]_{22}=( 2c^{n,j}_{k_3,1}\sqrt{3(n+1)}\bar{\gamma} \pm c^{n,j}_{k_3,2}\sqrt{n+2}(\gamma_1+6\bar{\gamma}))$, $[\v A^{n,j}_{k_3,\sigma}]_{33}=( c^{n,j}_{k_3,1}\sqrt{6}\bar{\gamma}k_3\lb \pm c^{n,j}_{k_3,3}\sqrt{n+1}(\gamma_1+6\bar{\gamma}))$, $[\v A^{n,j}_{k_3,\sigma}]_{44}=(-c^{n,j}_{k_3,2}\sqrt{6}\bar{\gamma}k_3\lb +2c^{n,j}_{k_3,3}\sqrt{3(n+2)}\bar{\gamma} \pm c^{n,j}_{k_3,4}\sqrt{n+3}(\gamma_1+4\bar{\gamma}))$, $[\v B^{n,j}_{k_3,\sigma}]_{11}=(c^{n,j}_{k_3,1}\sqrt{n+1}(\gamma_1+4\bar{\gamma})+2c^{n,j}_{k_3,2}\sqrt{3(n+2)}\bar{\gamma} +c^{n,j}_{k_3,3}\sqrt{6}\bar{\gamma}k_3\lb)$, $[\v B^{n,j}_{k_3,\sigma}]_{22} = ( c^{n,j}_{k_3,2}\sqrt{n+3}(\gamma_1+6\bar{\gamma}) - c^{n,j}_{k_3,4}\sqrt{6}\bar{\gamma}k_3\lb)$, $[\v B^{n,j}_{k_3,\sigma}]_{33} =(c^{n,j}_{k_3,3}\sqrt{n+2}(\gamma_1+6\bar{\gamma}) + 2c^{n,j}_{k_3,4}\sqrt{3(n+3)})$ and $[\v B^{n,j}_{k_3,\sigma}]_{44} =c^{n,j}_{k_3,4}\sqrt{n+4}(\gamma_1+4\bar{\gamma})$.
Here the upper(lower) sign of `$\pm$' corresponds to $\sigma=1(2)$.
As a result, the velocity matrix elements are non-vanishing only if the Landau level difference is one just like the case of the conventional two dimensional electron gas (2DEG).
Although the structure of the velocity operator of the Luttinger model is more complicated than that of the conventional 2DEG, one can note that they become essentially the same when $\bar{\gamma}$ vanishes.
This is because all the complexities originate from the inter-orbital mixing terms of the Hamiltonian (\ref{eq:hamiltonian_matrix}) and they are proportional to $\bar{\gamma}$.

\section{quantum oscillations}
For realistic evaluations of $\sigma_{11}$ and $\sigma_{12}$ for Pr$_2$Ir$_2$O$_7$, we may need more precise informations about the Luttinger parameters,
$\gamma_i$'s, $\kappa$ and $q$.
Although such informations are lacking at present, we notice that the qualitative structures of the Landau levels are robust for wide ranges of the parameters.
Thus we only fix the effective mass of the valence band to be six times the bare electron mass as a minimal consideration of the electronic structure,
which would be consistent with the previous tight-binding model constructions\cite{krempa}.
For this, we set the parameters as $\gamma_1 = 1$, $\bar{\gamma}=2$, $\kappa =1$, $q=-0.2$ and $m=6m_e$, where $m_e$ is the mass of the bare electron.
By converting the conductivities into the resistivities via $\rho_{11}=\sigma_{11}/(\sigma_{11}^2+\sigma_{12}^2)$ and $\rho_{12}=-\sigma_{12}/(\sigma_{11}^2+\sigma_{12}^2)$, we plot the magneto- and Hall resistivities as functions of the magnetic field in Fig. \ref{fig:transport}.
We consider both electron and hole doped cases with $n_e$ or $n_h = 1.4\times 10^{23}m^{-3}$.

The magneto-resistivitiy, starting from the finite zero-field resistivity, shows overall increasing trend due to the growing Landau level spacing as a function of the magnetic field.
%On the other hand, the Hall resistivity is vanishing at $B=0$ since we've neglected any sources which can break the time reversal symmetry such as the electron-electron interaction and the background magnetism.
The Hall resistivity $\rho_{12}$ exhibits the positive (negative) sign for the electron(hole)-doped case, where $\rho_{21}=-\rho_{12}$.
Notice that the Hall resistivity is vanishing at $B=0$ since we are considering a paramagnetic state.

The SdH signals are obtained from the magneto-resistivity curves by extracting their oscillating components.
The electron-doped case with $\Gamma=0.08$meV is plotted in Fig. \ref{fig:transport} (c). 
We checked that its local minima arise whenever the corresponding Fermi level touches one of the van-Hove singularities of the Landau level spectra.
When the magnetic field is decreased, the distance $\Delta (1/B)$ between two neighboring minima is suddenly widened above $B^{-1}\sim 0.2 \mathrm{T}^{-1}$.
The reason for this change in the periodicity is that the energy difference between two nearest neighboring Landau levels, such as $n=-2$ \& $n=0$ Landau levels near $E/\hbar\omega_c = 5$ and $n=-1$ \& $n=1$ near $E/\hbar\omega_c = 10$, become comparable to the Landau level broadening 
so that the broadened DOS cannot distinguish these Landau levels.

The quantum oscillation for the hole-doped case (Fig. \ref{fig:transport} (f)) shows more interesting features.
First, if $\Gamma=0.08$ meV, one can notice again that the local minima of the SdH curve are connected to the local maxima of the DOS in Fig.\ref{fig:band}(b).
However, when the impurity strength is weaken ($\Gamma=0.01$meV), one can observe more local minima than the previous one.
While this is also ascribed to the improved resolution by reducing the Landau level broadening, the minimum point $A_2$ is distinguishing in that it appears at the minimum point of the DOS as shown in Fig.\ref{fig:band}(b).
The origin of the new minimum in SdH signal can be found from the Fig.\ref{fig:band}(c), where we mark the corresponding Fermi levels by red dashed lines.
The minimum $A_2$ arises at the crossing point of two Landau levels with indices $n=-2$ and $n=-1$.
Since the difference between those two Landau level indices is one, the crossing point satisfies the selection rule of the velocity operator.
Furthermore, the transitions corresponding to $|\langle \psi_{k_3}^{n,j} | v_1| \psi_{k_3}^{m,i}\rangle|^2$ occur nearly or exactly at the same energies around the crossing points, hence the product of two Lorentzians in (\ref{eq:cond11}) gives a large contribution to the conductivity.

For both electron and hole doped cases, we can recover the periodic quantum oscillations in the (very) weak field limit.
On the other hand, the complicated SdH signals in the strong magnetic fields reflect the unconventional Landau level structures of the Luttinger model near the band touching point such as the inter-Landau level crossing and nonuniform Landau level spacings.

\section{Conclusions and Discussions}

The Luttinger model with quadratic band touching has been proposed as a minimal model for the electronic structure in the paramagnetic state of pyrochlore iridates.\cite{moon}
As mentioned earlier, a number of novel quantum ground states are shown to arise once appropriate perturbations are taken into account.\cite{yang,wan1,krempa,moon} Hence it is clearly important to establish the validity of the quadratic band touching spectra in this class of materials. We propose that quantum oscillations in magneto-transport in the paramagnetic states of pyrochlore iridates may present such an evidence via the aperiodic behaviors of SdH signals that come from unusual Landau level structures of the Luttinger model with quadratic band touching. In this work, we did not consider the effect of the long-range Coulomb interaction although it was shown, in a previous work where one of the authors was involved, that a non-Fermi liquid state arises when the electron-electron interaction is introduced to the Luttinger model with quadratic band touching. On the other hand, we expect that the qualitative features of the Landau-level structures and quantum oscillations would not change much as the renormalized dynamical exponent $z=1.9$ in three-dimensions
is not very different from $z=2$ of the quadratic band-touching spectra in the non-interacting model. Hence, for the purpose of our current work on the SdH signals, we believe the non-interacting model is a good starting point.

Indeed the aperiodic SdH signals\cite{onoda} observed in the paramagnetic states of Pr$_2$Ir$_2$O$_7$ are consistent with the theoretical results if a small hole-doping $n_h \sim 10^{23} m^{-3}$ and an effective mass of $m \sim 6 m_e$ are assumed for the quadratic band touching with moderate strength of impurity scattering ($\Gamma \sim 0.08 meV$). The range of high magnetic fields where the aperiodic SdH signals are seen and the estimated density of holes are consistent with the experimental findings.\cite{machida,nakatsuji,onoda} 

Combined with the recent ARPES experiments on Pr$_2$Ir$_2$O$_7$, where the signatures of the quadratic band touching is seen along the [111] direction\cite{nakatsuji14}, quantum oscillations may provide valuable informations about the underlying electronic structure that are responsible for the emergence of many novel quantum ground states. It will be very interesting to perform quantum oscillation experiments on other pyrochlore iridates and investigate the universality of the underlying theoretical model.

\acknowledgements

This work was supported by the NSERC of Canada, the Canadian Institute for Advanced Research, and Center for Quantum Materials at the University of Toronto (YBK).
We are grateful to S. Nakatsuji, T. Kondo, S. Shin, L. Balents, and E.-G.Moon for sharing their unpublished ARPES data on Pr$_2$Ir$_2$O$_7$ and discussions. J.R. thanks K. Hwang for useful discussions.

\end{document}